\documentclass{article}

\usepackage[utf8]{inputenc}
\pdfoutput=1
\usepackage{longtable}

\usepackage{booktabs}

\usepackage{amsmath}

\usepackage{url}

\usepackage{listings}

\usepackage{float}

\usepackage{lineno}

\usepackage{pgfplotstable}

\usepackage[left=3cm,right=3cm]{geometry}

\usepackage{etoolbox}
\AtBeginEnvironment{longtable}{%
   \setlength\thickmuskip{0mu}%
   \setlength\tabcolsep{3.7pt}
   \setlength\LTcapwidth{\textwidth}%
   }
   
\usepackage{pstricks-add}

\usepackage{tikz}

\usetikzlibrary{arrows.meta,backgrounds}

\usepackage[square,numbers]{natbib}

\providecommand{\keywords}[1]
{
  \small	
  \textbf{\textit{Keywords---}} #1
}

\title{Does home advantage without crowd exist in American football?}

\author{D\'{a}vid Zolt\'{a}n Szab\'{o}\footnote{Corvinus University of Budapest, Department of Finance.  E-mail: davidzoltan.szabo@uni-corvinus.hu.}, Diego Andr\'{e}s P\'{e}rez Ruiz\footnote{The University of Manchester, School of Social Statistics. E-mail: diego.perezruiz@manchester.ac.uk}}

\date{April 2021}

\begin{document}

\maketitle

\begin{abstract}
It is well-known that home team has an inherent advantage against visiting teams when playing team sports. One of the most obvious underlying reasons, the presence of supporting fans has mostly disappeared in major leagues with the emergence of COVID-19 pandemic. This paper investigates with the help of historical National Football League (NFL) data, how much effect spectators have on the game outcome. Our findings reveal that under no allowance of spectators the home teams' performance is substantially lower than under normal circumstances, even performing slightly worse than the visiting teams. On the other hand, when a limited amount of spectators are allowed to the game, the home teams' performance is no longer significantly different than what we observe with full stadiums. This suggests that from a psychological point of view the effect of crowd support is already induced by a fraction of regular fans.
\end{abstract}


\section{Introduction}


In professional sports leagues the games are regularly played either at home or at away venues. Home venue refers to the place where the given team is located and away venue to the place where the opposite team is located. There are occasional exceptions when the two teams play at neutral grounds that belong to neither of the two teams. Major sport championships are organized with a balanced schedule where each team plays around half of their matches at away or at home venues. American football is the most popular sport in the United States and the National Football League (NFL) seasons regularly had over 60,000 average number of visitors before the emergence of COVID-19 pandemic.

It is well established that teams have a sizeable advantage when playing at home court. Related studies with a detailed literature review on this notion can be found in \citep{carron2005home}, which provides a general overview of the causes and physiological implications, and \citep{pollard2005long}, which provides a review of the published research under the main explanations for home advantage in football. That said, leaving all other factors aside and assuming two equally strong teams, the team who plays at home venue has a higher chance of winning. There are a few potential causes as to why this phenomenon exists. For a review on the potential causes, see, e.g., \citep{nevill1999home, pollard2008home}. The possible mentioned causes are: crowd effects, travel effects, familiarity, referee bias, territoriality, special tactics, rule factors and psychological factors.

As pointed out in \citep{nevill1999home} and \citep{pollard2008home} these causes interact with each other and it is not necessarily possible to isolate these causes. Nonetheless, without the presence of supporting fans, two of the causes, namely the crowd effects and referee bias clearly vanish.
Regarding the other factors, visiting teams still have to spend time to reach the venue of the game, familiar conditions and a heightened sense of territoriality for the home team does not disappear without spectators, special tactics and rule factors are also not influenced by the presence of spectators. Finally the awareness of existence of home team advantage as a self perpetuating psychological factor for the players might be affected by the dearth of spectators as well. The lack of fans caused the players a state of confusion, whereby they were unable to sense the previous psychological advantage.

Considering association football (soccer) several studies \citep{nevill2002influence,garicano2005favoritism,dohmen2008influence,dawson2010influence} have shown evidence of referee bias by operating under an elevated social pressure and reacting in favour of the home team by awarding more penalties and a lower amount of disciplinary cards to the home team.

Distinguishing crowd support from other possible causes and examining to what extent crowd effect contribute to home advantage has been conducted in \citep{van2011supporters} and \citep{ponzo2018does}.
With the help of Italian ``Serie A" soccer league only those games were considered where the two teams share the same stadium. \citep{van2011supporters} surprisingly found that crowd support does not play any role in the observed home advantage, whereas \citep{ponzo2018does} argued that home teams scored significantly more goals and had a significantly higher chance of winning, even though the impact of home advantage is slightly lower than under normal games.

Due to the emergence of COVID-19 pandemic and the timing of National Football League (NFL) seasons, the 2019 NFL season could finish under normal circumstances in February 2020, whereas the 2020 NFL season has been impacted by the then ongoing COVID-19 pandemic \citep{Wiki2020NFL}.
Due to health and safety concerns no or only a greatly reduced amount of fans were allowed to be present at the stadiums during the games of the 2020 NFL season. Similarly, many other leagues, such as soccer leagues worldwide had to play without audience during the COVID-19 pandemic.
These unfortunate events enable the scientific analysis to assess to what extent home court advantage is underpinned by the presence of supporting fans. To the best of our knowledge there is no available study concerning NFL games without audience. On the other hand, published studies investigating the impact of COVID-19 pandemic on home advantage for soccer games are already available with conflicting results.
Examining German soccer games in 2020, \citep{tilp2020covid} found that COVID-19 lock-down led to a home disadvantage by observing more home losses than home wins. The authors of \citep{wunderlich2021does} found considering main European soccer leagues that only a non-significant decrease of home advantage was observed due to the COVID-19 pandemic, and \citep{matos2021home} found similar results considering Portuguese football League games.
\citep{bryson2021causal} studied refereeing decisions during COVID-19 pandemic and found that fewer cards were awarded to the away team, indicating a reduced social pressure on the referee including matches from twenty-three professional leagues and
seventeen countries in the 2019/20 season. \citep{winkelmann2021bookmakers} investigated the problem of bookmakers to adjust the betting odds after the emergence of COVID-19 pandemic including German soccer games and found that profitable betting strategies were therefore available.

The aim of this study is to investigate with rigorous statistical means whether for NFL games there is a significant difference between the 2020 NFL season and all previous seasons in terms of the performance of home and visiting teams. Our results show that home team's performance was strongly hindered by empty stadiums, resulting in an unprecedented lower than 50\% home win percentage without fans. Furthermore, we will also reveal that the allowance of some limited amount of fans recovered the performance of the home teams and they no longer substantially underperformed compared to historical results. 
These results can be interpreted that the psychological advantage derived from the awareness of supporting fans is already in effect with scatteredly filled stadiums. Moreover, we can claim that all other potential causes are irrelevant, the only real factors are crowd effects, referee bias and psychological factors. All these three factors are underlied by psychological concepts.

We conduct this research using game results along with their scores. Implementing statistical means, confidence intervals for the game results are estimated by binomial distributions, whereas difference of game scores of the teams are estimated by normal distributions. Moreover, classical statistical tests such as $Z$-tests and $t$-tests further support our findings. This study includes all seasons since 1970.

The rest of the paper is organised as follows. Section 2 discusses the considered data points for this analysis. Section 3 outlines the methodologies of this research. Section 4 presents the results and Section 5 concludes.

\section{Data}

We obtained NFL data from a few well established web resources \citep{profootballreference} and \citep{Shrpsports}. The website \url{http://www.shrpsports.com/} contains the historical NFL results for all games and for all seasons since 1933. The attendance data was extracted from \url{https://www.pro-football-reference.com/years/2020/attendance.htm}. Each downloaded table was checked for inconsistencies and saved into a database. 

Using these websites we can get sufficient information to conduct an analysis on home advantage. We gathered the site of the game along with the scores of the home and visiting teams for all games played in the examined NFL seasons. For each season, we did not take into account the games that were played at neutral sites and the games that ended with a draw and so we precluded these games. These games are indeed one-off events, in each season only up to a few games are tie or are not played at either team's site, besides games at neutral sites are irrelevant to this study where we intend to investigate home court advantage.
For the sake of this study, we considered the time period 1970-2020. This covers 51 NFL seasons, even though we have data from earlier time periods, 1970 marks the first season after the American Football League (AFL) and the NFL merged \citep{felser2008birth}. This merger has transformed the football significantly, and we chose this year for a cut-off year to start the investigation of this study.

As our main goal is to show that last season's results are significantly different, we draw particular attention to the available data from Season 2020. Games of Season 2020 can be further distinguished based on the number of fans attending a particular game. Indeed there were 144 games played without audience and 122 games played with limited audience in Season 2020. This relatively even distribution provides us further means to examine the difference of events when there is absolutely no audience or when the audience is only limited.

\section{Methods}\label{sect:Methods}

Here we discuss the methodologies that were used to complete this research. We treat the analysis on the match outcome and on the number of points achieved by the two teams separately.

Given our big database we can calculate results separately for games within a given season. As for Season 2020, we make a further separation to calculate results for three different scenarios: 
\begin{enumerate}
    \item[] \textit{Scenario 1}: Season 2020 with no audience
    \item[] \textit{Scenario 2}: Season 2020 with limited audience
    \item[] \textit{Scenario 3}: Season 2020 both with no or limited audience
\end{enumerate}
Under \textit{Scenario 1} we consider games from Season 2020 when absolutely no fans were allowed to the stadium, under \textit{Scenario 2} we consider games from Season 2020 when some limited amount of fans were allowed to the stadium, and under \textit{Scenario 3} we consider all games from Season 2020. We also combine observations from Season 1970 until Season 2019 into one sample, refer to it as \textit{merged 1970-2019 sample}, and calculate results for this merged sample as well.


\subsection{Analysis of game outcome}\label{Methods:1}

First, we mathematically formulate the sample proportion (denoted by $\hat{\pi}$). Let

\begin{equation}\label{eq:1}
\hat{\pi}=\frac{\texttt{\# Wins by home team}}{\texttt{\# Wins by home team}+\texttt{\# Wins by away team}}    
\end{equation}

be the proportion of wins by the home team using a certain sample. This sample proportion plays a crucial role in this study. 

Once $\hat{\pi}$ have been calculated for different samples, we turn to a rigorous statistical analysis including the calculation of confidence intervals. Assuming that the true likelihood of home team wins is constant $\pi$ for a particular sample (leaving all other factors aside, or assuming that both teams are equally strong), then the number of home team wins for a certain sample follows a binomial distribution, $Bin(n,\pi)$, with parameters $n$ and $\pi$, where $n$ is the number of games within a sample. This assumption is reasonable, as the number of games is sufficiently large and each team is supposed to play around half of their matches at home in a given season. Therefore the possible effect that the home team is significantly weaker or stronger than the away team cancels out when considering an entire season together.

The corresponding confidence intervals for $\pi$ will be chosen as binomial proportion confidence intervals at 95\% and 99\% confidence levels using normal approximations.
The following formula quantifies this:


\begin{equation}\label{bin:confinterv}
\hat{\pi} \pm \underbrace{ \text{Z}^{*} \cdot \sqrt{\frac{\hat{\pi}\cdot (1-\hat{\pi})}{n}} }_\text{M} 
 \end{equation}

where $\text{Z}^{*}$ is the $1-\frac{\alpha}{2}$ quantile of the standard normal distribution corresponding to the $\alpha$ confidence level and $\text{M}$ is the margin of error.
The value of $\text{Z}^{*}$ is 1.96 for 95\% confidence level and 2.58 for 99\% confidence level.

We will also use the classical two sample $Z$-test to test the following null hypothesis:

\begin{equation*}
    H_0: \pi_1=\pi_2, \quad \text{against} \quad  H_A : \pi_1 \neq \pi_2,
\end{equation*}

where $\pi_1$ and $\pi_2$ are the true success probabilities to sample one and sample two, respectively.

To perform this, the following test statistic will be used:

\begin{equation}
    z=\frac{\hat{\pi_1}-\hat{\pi_2}}{\sqrt{\frac{\hat{\pi_1}(1-\hat{\pi_1})}{n_1}+\frac{\hat{\pi_2}(1-\hat{\pi_2})}{n_2}}},
\end{equation}

where $\hat{\pi_1}$ and $\hat{\pi_2}$ are the observed sample proportions to sample one and sample two, respectively.
The value of $z$ will be compared to $\text{Z}^{*}$ at 95\% or 99\% confidence levels.

With the help of confidence intervals and two sample $Z$-tests we are able to compare different years and perform a comprehensive analysis of the research question.

\subsection{Analysis of scores by teams}\label{Method:Scores}

As for our second analysis, we investigate the number of points achieved by home and away teams and the difference between these scores.

The mean of the scores can be calculated for each season in a straightforward way. Nonetheless, determining the distribution of the scores is not straightforward contrary to the problem we faced in Section~\ref{Methods:1}. 
Our primary aim with fitting distributions is to create a confidence interval for the mean of the scores. 
We invoke previous studies to better cope with this problem. 
\citep{pollard1973collegiate} claimed that points in collegiate football games exhibit negative binomial properties. As for other team sports, in \citep{gill2000late} the authors used Poisson distribution to model Hockey scores due to it being a low scoring sport. Other low scoring sports points such as soccer has also been effectively modelled using Poisson distributions \citep{dixon1997modelling,lee1997modeling}, which is not the case for NFL.
In \citep{gill2000late} the authors used a normal distribution approach to model the difference between home and away scores for NFL games following the findings of previous related article \citep{stern1991probability}. 

Let $H$ and $A$ denote, respectively, the score of the home and away team at the end of the match. 
We will infer in Section~\ref{Results} that for this current study we only need to model the difference of two team's points ($\text{D}= H - A$). As this difference can take both negative and positive values we preclude the Poisson and negative binomial distributions from consideration.
Hence, we will assume that $\text{D}$ follows a normal distribution, say $N(\mu,\sigma^2)$ for a given sample. Even though the scores are integer numbers contrary to normally distributed random variables that can take any non-integer real values as well, a relatively accurate approximation can still be provided using normal distributions. Regarding our observations, we will show that the assumption of normality for $\text{D}$ is reasonable in Section~\ref{Results}.



We are also able to estimate confidence intervals for $\mu$ and $\sigma$ parameters using the ``fitdist" and ``confint" \textsf{R} functions that are contained in package ``fitdistrplus" \citep{fitdistrplus} and ``stats" \citep{stats}, respectively. Based on these confidence intervals we can compare the scores achieved by the teams for the different samples and conclude.

We will also use the classical two sample $t$-test with unequal sample sizes (Welch's $t$-test) to test the following null hypothesis:
\begin{equation}
    H_0: \mu_1=\mu_2 \quad \text{against} \quad H_A: \mu_1 \neq\mu_2,
\end{equation}
where $\mu_1$ and $\mu_2$ are the true means of score differences to sample one and sample two, respectively.

To perform this the following test statistic will be used:

\begin{equation}
    T=\frac{\bar{Y_1}-\bar{Y_2}}{\sqrt{\frac{s_1^2}{n_1}+\frac{s_2^2}{n_2}}},
\end{equation}
where $\bar{Y_1}$ and $\bar{Y_2}$ are the sample means, and $s_1^2$ and $s_2^2$ are the sample variances.
The value of $T$ will be compared to $\text{t}_{1-\frac{\alpha}{2},\nu}$, that is the critical value of the $t$ distribution with $\nu$ degrees of freedom.

The following formula gives the degrees of freedom
\begin{equation}
\nu=\frac{\left(\frac{s_1^2}{n_1} + \frac{s_2^2}{n_2}\right)^2}{\frac{\left(s_1^2/n_1\right)^2}{n_1-1} + \frac{\left(s_2^2/n_2\right)^2}{n_2-1}}.
\end{equation}

We will again use $\alpha=95\%$ or $\alpha=99\%$ confidence intervals and we can reject the null hypothesis, in case $|T|>\text{t}_{1-\frac{\alpha}{2},\nu}$. With the help of confidence intervals and two sample $t$-tests we are again able to make a comparison and decide if the mean of score difference is substantially different or not over the seasons.

\section{Results}\label{Results}

In this section we will discuss the results corresponding to the research problem outlined in Section \ref{sect:Methods}.

\subsection{Results on game outcome}

\begin{figure}[H]
  \includegraphics[width=.45\textwidth]{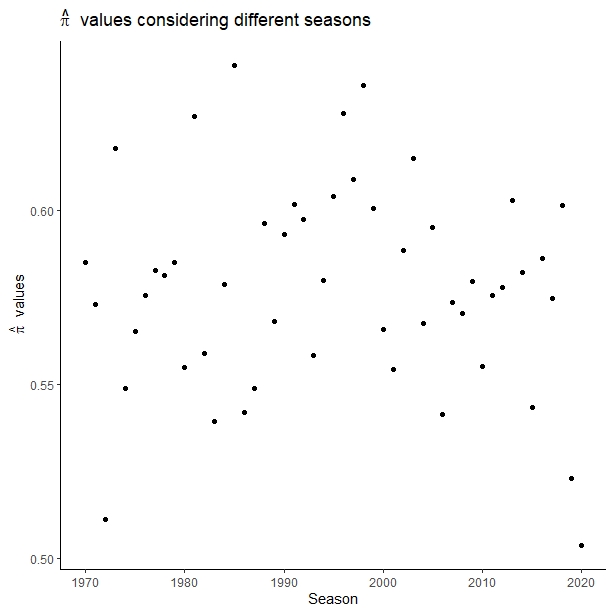}
   \hspace{1cm}
  \includegraphics[width=.45\textwidth]{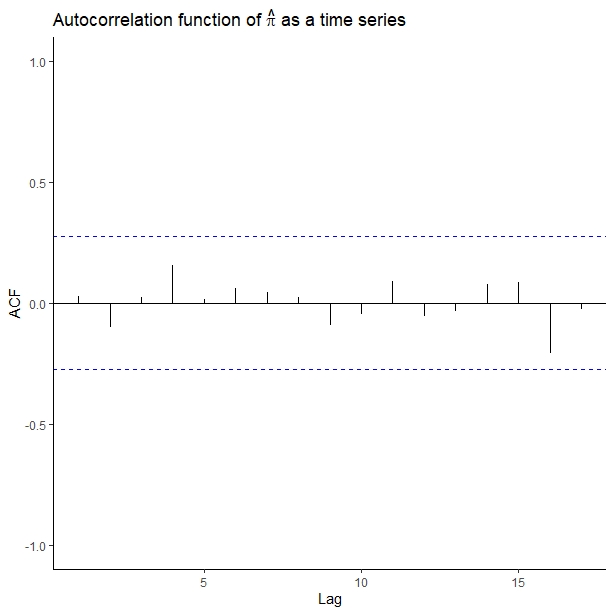}
  \caption{The left graph shows the values of $\pi$ for different seasons, whereas the right graph shows the autocorrelation function of the values of $\pi$.}
  \label{Fig_Prop}
\end{figure}

\begin{figure}[H]
  \includegraphics[width=1\textwidth]{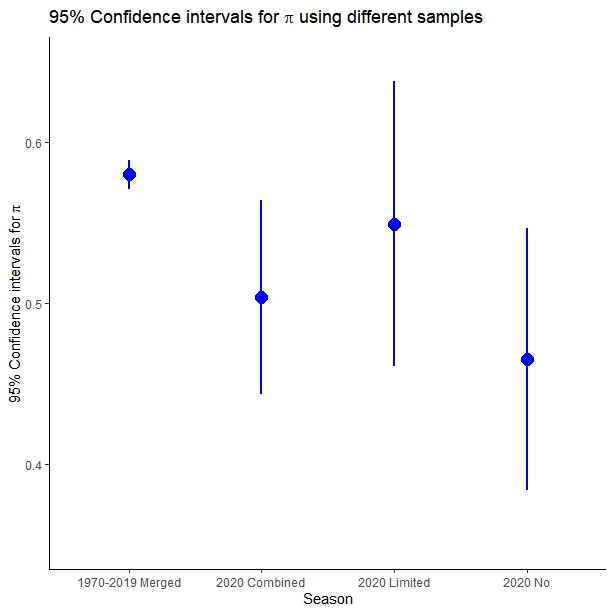}
  \caption{This figure shows the 95\% confidence intervals of parameter $\pi$ considering different samples. From the left to right, we can see the corresponding confidence intervals, where all games from the \textit{merged 1970-2019 sample} were considered, all games from Season 2020 were considered, games from Season 2020 with limited attendance were considered, and games from Season 2020 with strictly no spectators were considered.}
  \label{Fig_ConfInt}
\end{figure}

We gather the most important results into Table \ref{Table:confintervals}. This table shows the number of games played for the different samples, also the value of $\hat{\pi}$'s along with the 95\% and 99\% confidence intervals for $\pi$. 
Figure~\ref{Fig_Prop} plots the realised $\hat{\pi}$ values against the 51 different seasons along with its autocorrelation function, whereas Figure~\ref{Fig_ConfInt} shows the 95\% confidence interval of $\pi$ for the \textit{merged 1970-2019 sample}, and for the three scenarios of Season 2020.


Equation~\ref{bin:confinterv} reveals the width (double $\text{M}$) of the confidence interval, that depends on the number of observations and on the chosen confidence level. The number of games in an NFL season has grown a couple of times, still the considered games for each season are between 176 and 266, with the exception of Season 1982 when due to an 57-day-long players' strike the number of games were reduced \citep{lock1985national}. As for the last season, a total of 144 games were played without absolutely any fans, and 122 games were played with limited audience \citep{profootballreference}, therefore the sample size under the three scenarios of Season 2020 are 144, 122 and 266. The actual limitations for Season 2020 varied among the home sites, but none of the teams allowed spectators at more than 25\% capacity with the exception of the Super Bowl \citep{Wiki2020NFL}. This resulted in the average attendance decline from 66151 to 5721 between 2019 and 2020 NFL season \citep{StatistaNFL}. 

As we can see in Table~\ref{Table:confintervals} and in Figure~\ref{Fig_Prop}, the value of $\hat{\pi}$ was higher than 50\% for all 51 entire seasons, strongly indicating the existence of the notion of home advantage in general ($\pi>50\%$). 
The value of $\hat{\pi}$ apart from the 2020 Season was the lowest during the 1972 NFL season equalling 51.11\%. Considering all games in 2020 Season combined, the value of $\hat{\pi}$ is 50.38\% being the lowest among all considered seasons.
Moreover, considering the games with no audience this $\hat{\pi}$ value dropped down to 46.53\%, being the only $\hat{\pi}$ value in Table~\ref{Table:confintervals} below 50\%. 
These $\hat{\pi}$ values already suggest that the 2020 NFL season had a significantly lower amount of home wins than the previous seasons.
We aim to confirm this claim and also the existence of general home advantage with spectators by comparison of confidence intervals for $\pi$.

In Table \ref{Table:confintervals} we also show corresponding results to the \textit{merged 1970-2019 sample}.
The corresponding confidence interval of $\pi$ is very narrow in comparison to other confidence intervals due to the large number (11765) of considered games.
For 21 different seasons even the 99\% confidence intervals are strictly above 50\% meaning that for these 21 seasons we can immediately reject the hypothesis that home advantage does not exist, or in other words that $\pi \le 50\%$. 
Furthermore, the 99\% confidence interval considering the \textit{merged 1970-2019 sample} is [56.81\%,59.16\%]. This shows that with 99\% probability the true probability of a home team win with fans at site is within this interval. This is reassuring of the existence of home court advantage in general.

\begin{figure}[hbt!]
\centering

\begin{tikzpicture}
\draw [-latex,very thick] (0,0) -- (6,0);
\draw [Bar-Bar,very thick] (1,2) -- node[fill,circle] (a) {} +(2,0);
\draw [Bar-Bar,very thick] (3.5,1) -- node[fill,circle] (b) {} +(2,0);
\draw [densely dotted] (a) -- (a|-0,0)
                       (b) -- (b|-0,0);
\node at (2,-0.5) {$\pi_1$};
\node at (4.5,-0.5) {$\pi_2$};
\end{tikzpicture}
\caption{Visual representation of non-overlapping confidence intervals for $\pi_1$ and $\pi_2$ in the case when $\pi_1 < \pi_2$.} \label{fig:diagram}
\end{figure}
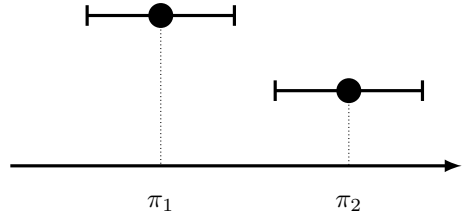

As our goal is to distinctively examine the 2020 NFL season, we now pay particular attention to the corresponding confidence intervals of Season 2020.
In case there is no overlap between confidence intervals we can claim that the $\pi$ parameter values expressing the true chance of home win are significantly different.
This is because even using 95\% confidence intervals the probability that the true value of parameter $\pi$ is not contained in the interval is 5\%, and the probability that the true $\pi$ value is above or below the interval is $(2.5\%)\cdot(2.5\%)$. If the 95\% confidence intervals for $\pi_1$ and $\pi_2$ do not overlap in the sense that the confidence interval for $\pi_2$ is strictly above the confidence interval of $\pi_1$, then the probability that $\pi_1<\pi_2$ is at least $(1-2.5\%\cdot 2.5\%)=99.9375\%$. See Figure~\ref{fig:diagram} for a visual interpretation. Therefore we argue that with non overlapping 95\% confidence intervals we can come to the conclusion that one of the $\pi$ values is significantly higher than the other one.


Figure~\ref{Fig_ConfInt} shows the three corresponding 95\% level confidence intervals for $\pi$ to Season 2020 along with the confidence interval for $\pi$ to the \textit{merged 1970-2019 sample}. 
Considering the scenario with limited audience of Season 2020 the confidence interval does overlap with the confidence interval of the \textit{merged 1970-2019 sample}. On the other hand, there is no overlap between the confidence interval of the \textit{merged 1970-2019 sample} with the confidence interval of the scenario of Season 2020 with no audience, and with the confidence interval of the scenario of Season 2020 both with no or limited audience.
Thus, we can conclude that for Season 2020 the true probability of home court win was significantly lower than for previous seasons under normal circumstances. Besides, we cannot conclude the same when considering the subset of games when some limited amount of fans were at the stadium. Indeed, the corresponding 54.92\% value of $\hat{\pi}$ is not an outlier when inspecting all $\hat{\pi}$ values in Table~\ref{Table:confintervals}.
Finally, even though the value of $\hat{\pi}$ of Season 2020 with no fans is below 50\%, yet the corresponding confidence interval for $\pi$ covers the 50\% value. Therefore we cannot conclude that true home win chance was significantly lower than 50\% without spectators, only that true home win likelihood was significantly lower than what we have seen for previous seasons. 

Regarding the two sample $z$-tests statistics, we compared different samples to the \textit{merged 1970-2019 sample}. As for the actual $z$ statistic values, we obtain $z=2.46$ with \textit{Scenario 1} of Season 2020, $z=0.68$ with \textit{Scenario 2} of Season 2020 and $z=2.74$ with \textit{Scenario 3} of Season 2020. Therefore at 95\% confidence level we can reject that the true success probability of the \textit{merged 1970-2019 sample} is the same as for the sample containing all games of Season 2020; and that the true success probability of the \textit{merged 1970-2019 sample} is the same as for the sample of games of Season 2020 with no audience. We can reject the null hypothesis that the true success probability of the \textit{merged 1970-2019 sample} is the same as the sample of games of Season 2020 with no audience even at $99\%$ confidence level. Regarding the games of Season 2020 with some limited audience we cannot make the likewise rejection even at 95\% confidence level.

We also calculated $z$-test statistics values as a comparison to the \textit{merged 1970-2019 sample} by choosing those seasons from Table \ref{Table:confintervals} that have the lowest or highest $\hat{\pi}$ values other than Season 2020. We obtain $z=1.83$ for Season 1972, $z=-1.94$ for Season 1985 and $z=1.81$ for Season 2019. Therefore even at $95\%$ confidence level we cannot reject the null hypothesis that the real success probabilities of the two samples are the same.

To sum up, based on either the confidence interval comparison or on the $z$-test statistics we can conclude that the only true outlier season in terms of home team's success probability was Season 2020, and the outlier nature of Season 2020 is even stronger only considering those observations when there were no spectators at the stadium.

\begin{center}
\begin{longtable}{|c|c|c|c|c|c|c|c|}
\caption{This table shows yearly statistics of home advantage. For each season we can see the total number of games played, how many out of this were won and lost by the home team, how big the values of $\hat{\pi}$-s were. Besides, we can also see 95\% and 99\% confidence intervals for $\pi$ using the binomial distribution approach. We also calculated these numbers merging data from Seasons 1970 until 2019 into one sample. For the last season we considered the analysis for the three scenarios.}\\
\hline
\textbf{Season} & \textbf{Fans?}& \textbf{$n$} & \textbf{Won}& \textbf{Lost} & \textbf{$\hat{\pi}$}& \textbf{95\% C.I. for $\pi$} & \textbf{99\% C.I. for $\pi$} \\
\hline
\endfirsthead
\multicolumn{8}{c}%
{\tablename\ \thetable\ -- \textit{Continued from previous page}} \\
\hline
\textbf{Season} & \textbf{Fans?}& \textbf{$n$} & \textbf{Won}& \textbf{Lost} & \textbf{$\hat{\pi}$}& \textbf{95\% C.I. for $\pi$} & \textbf{99\% C.I. for $\pi$} \\
\hline
\endhead
\hline \multicolumn{4}{r}{\textit{Continued on next page}} \\
\endfoot
\hline
\endlastfoot
1970 & Yes& 176 & 103 & 73 & 58.52\% & [51.24\%,65.80\%] & [48.94\%,68.10\%] \\

1971 & Yes& 178 & 102 & 76 & 57.30\% & [50.04\%,64.57\%] & [47.74\%,66.87\%] \\

1972 & Yes& 180 & 92 & 88 & 51.11\% & [43.81\%,58.41\%] & [41.50\%,60.72\%] \\

1973 & Yes& 178 & 110 & 68 & 61.80\% & [54.66\%,68.94\%] & [52.40\%,71.19\%] \\

1974 & Yes& 184 & 101 & 83 & 54.89\% & [47.70\%,62.08\%] & [45.43\%,64.36\%] \\

1975 & Yes& 184 & 104 & 80 & 56.52\% & [49.36\%,63.68\%] & [47.09\%,65.95\%] \\

1976 & Yes& 198 & 114 & 84 & 57.58\% & [50.69\%,64.46\%] & [48.51\%,66.64\%] \\

1977 & Yes& 199 & 116 & 83 & 58.29\% & [51.44\%,65.14\%] & [49.27\%,67.31\%] \\

1978 & Yes& 227 & 132 & 95 & 58.15\% & [51.73\%,64.57\%] & [49.70\%,66.60\%] \\

1979 & Yes& 229 & 134 & 95 & 58.52\% & [52.13\%,64.90\%] & [50.12\%,66.92\%] \\

1980 & Yes& 227 & 126 & 101 & 55.51\% & [49.04\%,61.97\%] & [47.00\%,64.02\%] \\

1981 & Yes& 228 & 143 & 85 & 62.72\% & [56.44\%,69.00\%] & [54.46\%,70.98\%] \\

1982 & Yes& 136 & 76 & 55 & 55.88\% & [47.54\%,64.23\%] & [44.90\%,66.87\%] \\

1983 & Yes& 228 & 123 & 105 & 53.95\% & [47.48\%,60.42\%] & [45.43\%,62.46\%] \\

1984 & Yes& 228 & 132 & 96 & 57.89\% & [51.49\%,64.30\%] & [49.46\%,66.33\%] \\

1985 & Yes& 229 & 147 & 82 & 64.19\% & [57.98\%,70.40\%] & [56.02\%,72.37\%] \\

1986 & Yes& 227 & 123 & 104 & 54.19\% & [47.70\%,60.67\%] & [45.65\%,62.72\%] \\

1987 & Yes& 215 & 118 & 107 & 54.88\% & [48.23\%,61.54\%] & [46.13\%,63.64\%] \\

1988 & Yes& 228 & 136 & 92 & 59.65\% & [53.28\%,66.02\%] & [51.27\%,68.03\%] \\

1989 & Yes& 227 & 129 & 98 & 56.83\% & [50.38\%,63.27\%] & [48.35\%,65.31\%] \\

1990 & Yes& 231 & 137 & 94 & 59.31\% & [52.97\%,65.64\%] & [50.97\%,67.65\%] \\

1991 & Yes& 231 & 139 & 92 & 60.17\% & [53.86\%,66.49\%] & [51.86\%,68.48\%] \\

1992 & Yes& 231 & 138 & 93 & 59.74\% & [53.42\%,66.06\%] & [51.42\%,68.07\%] \\

1993 & Yes& 231 & 129 & 102 & 55.84\% & [49.44\%,62.25\%] & [47.41\%,64.27\%] \\

1994 & Yes& 231 & 134 & 97 & 58.01\% & [51.64\%,64.37\%] & [49.63\%,66.39\%] \\

1995 & Yes& 250 & 151 & 99 & 60.40\% & [54.34\%,66.46\%] & [52.42\%,68.38\%] \\

1996 & Yes& 250 & 157 & 93 & 62.80\% & [56.81\%,68.79\%] & [54.91\%,70.69\%] \\

1997 & Yes& 248 & 151 & 97 & 60.89\% & [54.81\%,66.96\%] & [52.89\%,68.88\%] \\

1998 & Yes& 250 & 159 & 91 & 63.60\% & [57.64\%,69.56\%] & [55.75\%,71.45\%] \\

1999 & Yes& 258 & 155 & 103 & 60.08\% & [54.10\%,66.05\%] & [52.21\%,67.94\%] \\

2000 & Yes& 258 & 146 & 112 & 56.59\% & [50.54\%,62.64\%] & [48.63\%,64.55\%] \\

2001 & Yes& 258 & 143 & 115 & 55.43\% & [49.36\%,61.49\%] & [47.44\%,63.41\%] \\

2002 & Yes& 265 & 156 & 109 & 58.87\% & [52.94\%,64.79\%] & [51.07\%,66.67\%] \\

2003 & Yes& 265 & 163 & 102 & 61.51\% & [55.65\%,67.37\%] & [53.80\%,69.22\%] \\

2004 & Yes& 266 & 151 & 115 & 56.77\% & [50.81\%,62.72\%] & [48.93\%,64.60\%] \\

2005 & Yes& 257 & 153 & 104 & 59.53\% & [53.53\%,65.53\%] & [51.63\%,67.43\%] \\

2006 & Yes& 266 & 144 & 122 & 54.14\% & [48.15\%,60.12\%] & [46.25\%,62.02\%] \\

2007 & Yes& 265 & 152 & 113 & 57.36\% & [51.40\%,63.31\%] & [49.52\%,65.20\%] \\

2008 & Yes& 263 & 150 & 113 & 57.03\% & [51.05\%,63.02\%] & [49.16\%,64.91\%] \\

2009 & Yes& 264 & 153 & 111 & 57.95\% & [52.00\%,63.91\%] & 
[50.12\%,65.79\%] \\

2010 & Yes& 263 & 146 & 117 & 55.51\% & [49.51\%,61.52\%] & [47.61\%,63.42\%] \\

2011 & Yes& 264 & 152 & 112 & 57.58\% & [51.61\%,63.54\%] & [49.73\%,65.42\%] \\

2012 & Yes& 263 & 152 & 111 & 57.79\% & [51.83\%,63.76\%] & [49.94\%,65.65\%] \\

2013 & Yes& 262 & 158 & 104 & 60.31\% & [54.38\%,66.23\%] & [52.51\%,68.10\%] \\

2014 & Yes& 261 & 152 & 109 & 58.24\% & [52.25\%,64.22\%] & [50.36\%,66.11\%] \\

2015 & Yes& 265 & 144 & 121 & 54.34\% & [48.34\%,60.34\%] & [46.45\%,62.23\%] \\

2016 & Yes& 261 & 153 & 108 & 58.62\% & [52.65\%,64.60\%] & [50.76\%,66.49\%] \\

2017 & Yes& 261 & 150 & 111 & 57.47\% & [51.47\%,63.47\%] & [49.58\%,65.37\%] \\

2018 & Yes& 261 & 157 & 104 & 60.15\% & [54.21\%,66.09\%] & [52.33\%,67.97\%] \\

2019 & Yes& 260 & 136 & 130 & 52.31\% & [46.24\%,58.38\%] & [44.32\%,60.30\%] \\

1970-2019 Merged& Yes& 11765 & 6822 & 4943 & 57.99\% & [57.09\%,58.88\%] & [56.81\%,59.16\%] \\

2020 & No& 144 & 67 & 77 & 46.53\% & [41.57\%,54.67\%] & [35.80\%,57.26\%] \\

2020 & Limited& 122 & 67 & 55& 54.92\% & [46.09\%,63.75\%] & [43.30\%,66.54\%] \\

2020 & Combined & 266 & 134 & 132& 50.38\% & [44.37\%,56.38\%] & [42.47\%,58.29\%] \\
\end{longtable}
\label{Table:confintervals}
\end{center}

\subsection{Results on game scores}

Here we discuss the results of our study on the number of points achieved by both teams ($H$ and $A$) and the difference ($D$) between these points.

\begin{figure}[H]
  \includegraphics[width=.45\textwidth]{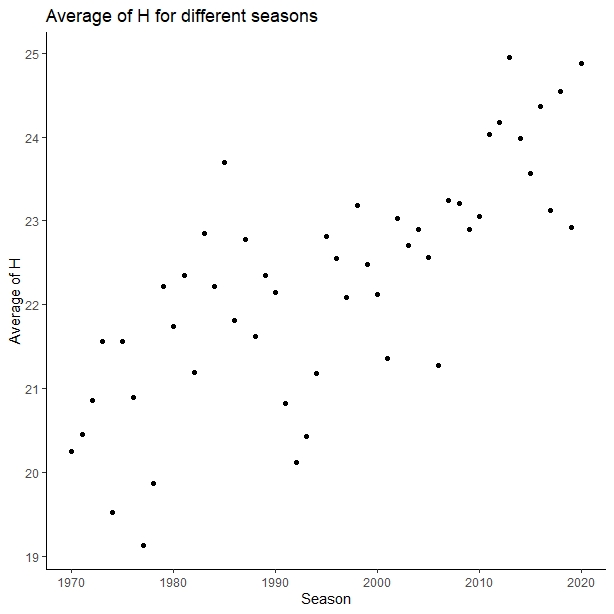}
   \hspace{1cm}
  \includegraphics[width=.45\textwidth]{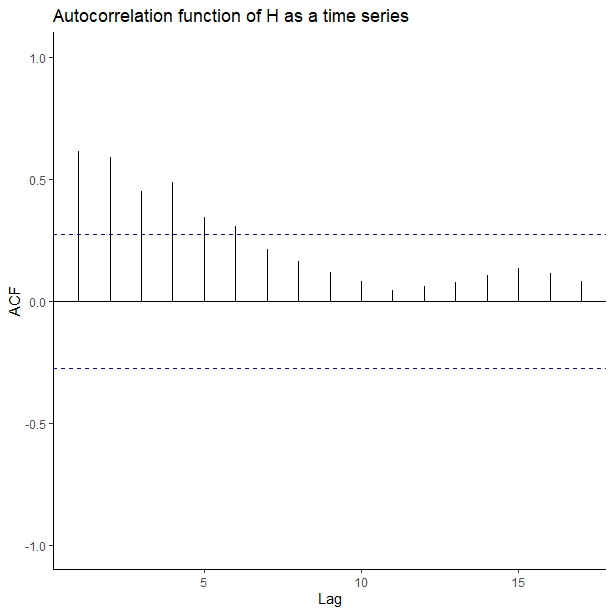}
  \caption{The left graph shows the average of home team's points for different seasons, whereas the right graph shows the autocorrelation function of the yearly average of home team's points.}
  \label{Fig_Home}
\end{figure}

\begin{figure}[H]
  \includegraphics[width=.45\textwidth]{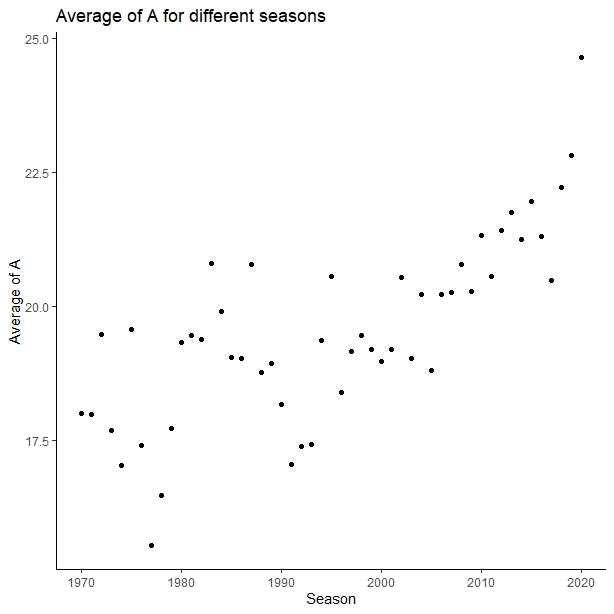}
   \hspace{1cm}
  \includegraphics[width=.45\textwidth]{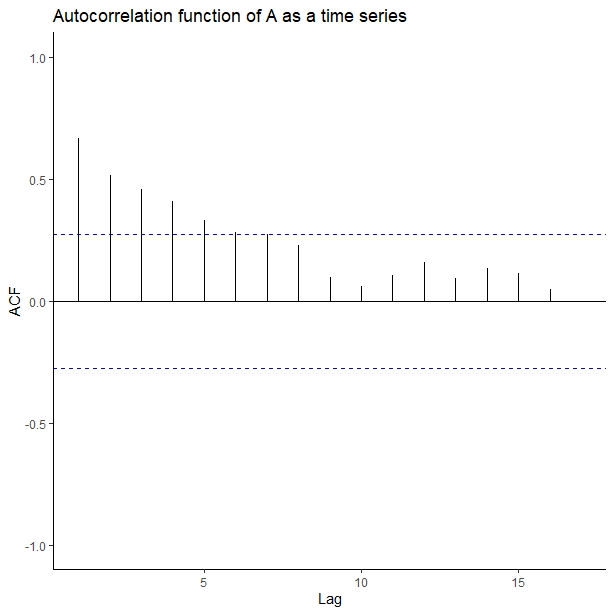}
  \caption{The left graph shows the average of away team's points for different seasons, whereas the right graph shows the autocorrelation function of the yearly average of away team's points.}
  \label{Fig_Away}
\end{figure}

On the left of Figures \ref{Fig_Home}--\ref{Fig_Away} we can see the average of $H$ and $A$ for different seasons, respectively. Inspecting these graphs more closely one can observe slight increasing trends, meaning that over the recent seasons both teams tended to score more point on average than during earlier seasons.
On the other hand, we can also see that the average $H$ is the second highest for Season 2020 and the average of $A$ is outstandingly the highest for Season 2020. In any case, the right graphs of Figures \ref{Fig_Home}--\ref{Fig_Away} reveal that for both $H$ and $A$ there is a significant level of autocorrelation, if we consider them as time series. This further supports the increasing trend phenomenon of $H$ and $A$. Therefore observing high $H$ and $A$ values for Season 2020 might also be the result of the increasing nature of this time series and it would not necessarily reflect the outstanding nature of Season 2020. Thus, we henceforth consider $D$.

\begin{figure}[H]
  \includegraphics[width=.45\textwidth]{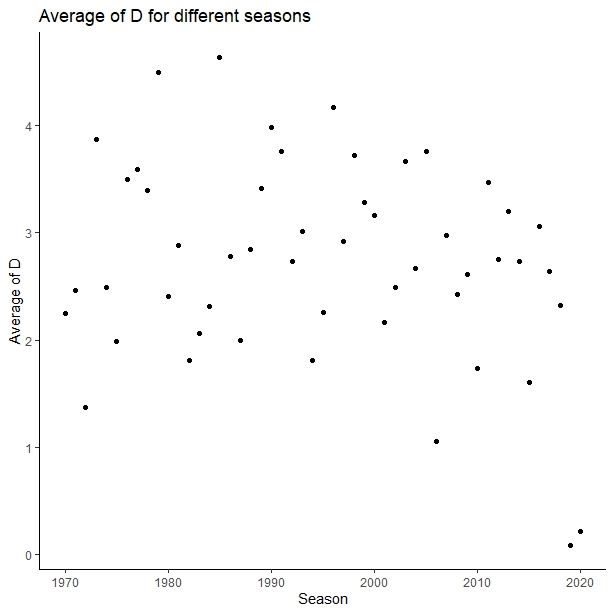}
   \hspace{1cm}
  \includegraphics[width=.45\textwidth]{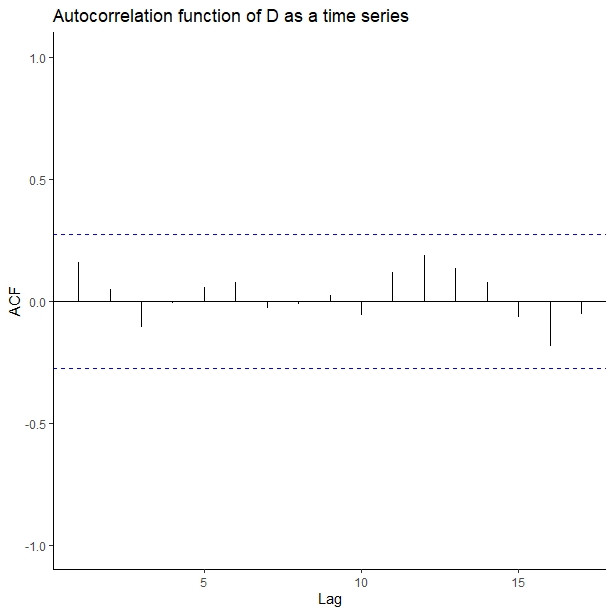}
  \caption{The left graph shows the average of the difference of home team's and away team's points for different seasons, whereas the right graph shows the autocorrelation function of the yearly average of the difference of home team's and away team's.}
  \label{Fig_Diff}
\end{figure}

On the left graph of Figure~\ref{Fig_Diff} we can see the average of $D$ for different seasons. Opposed to the season average of home and away points, we cannot observe a clear increasing or decreasing trend in this graph. As an interesting outlier, Season 2019 has the lowest average of $D$, with a value even being slightly lower than that of Season 2020.
On the right graph of Figure~\ref{Fig_Diff} we can see the corresponding autocorrelation function of the average of $D$ considered as a time series which reveals that this time series does not exhibit a significant level of autotocorrelation, providing a confirmation to the claim that the average of $D$ as a time series does not contain any trends. 

\begin{figure}[H]
  \includegraphics[width=.45\textwidth]{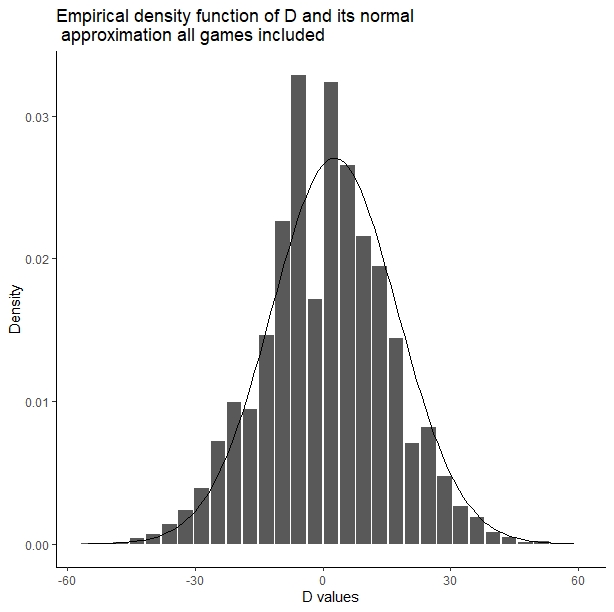}
   \hspace{1cm}
  \includegraphics[width=.45\textwidth]{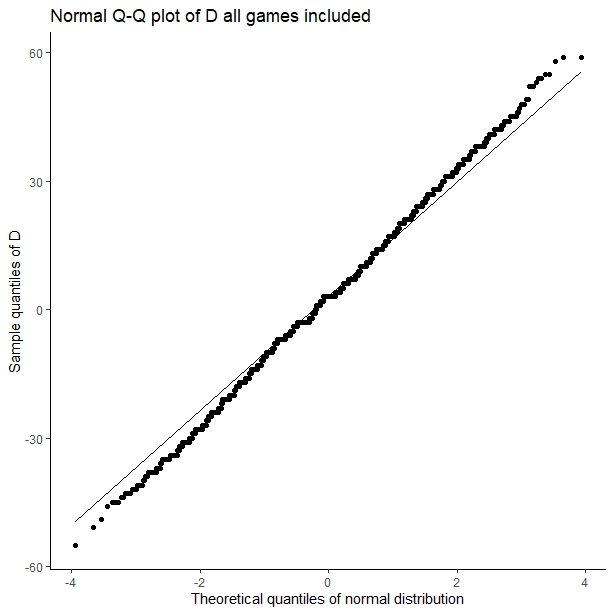}
  \caption{The left graph shows the empirical density of the difference of home team's and away team's points considering all games between Season 1970 and 2020, whereas the right graph shows the Normal Q-Q plot of the difference of home team's and away team's points considering all games between Season 1970 and 2020.}
  \label{Fig_HistQQ}
\end{figure}

As discussed in Section~\ref{Method:Scores}, supported by the related literature it is practical to fit normal distribution for $D$.
We indeed fit the a normal distribution for $D$ including all games in Seasons 1970-2020 and then separately for the games belonging to the 51 individual seasons, for the three scenarios of Season 2020 and also for the \textit{merged 1970-2019 sample}.
Regarding the goodness of the fit, Shapiro–Wilk test was conducted for the 51 individual seasons' fit and this test gives greater than 0.05 p-value for 42 fits.
Moreover, Figure \ref{Fig_HistQQ} compares the empirical density function to the normal density function and also shows the normal Q-Q plot when including all data points from Seasons 1970-2020.
Figure~\ref{Fig_HistQQ} provides a good indication of normality.
This along with Shapiro–Wilk tests, confirms that normal distribution is a very reasonable choice for modelling the difference of the number of points.

\begin{figure}[H]
  \includegraphics[width=1\textwidth]{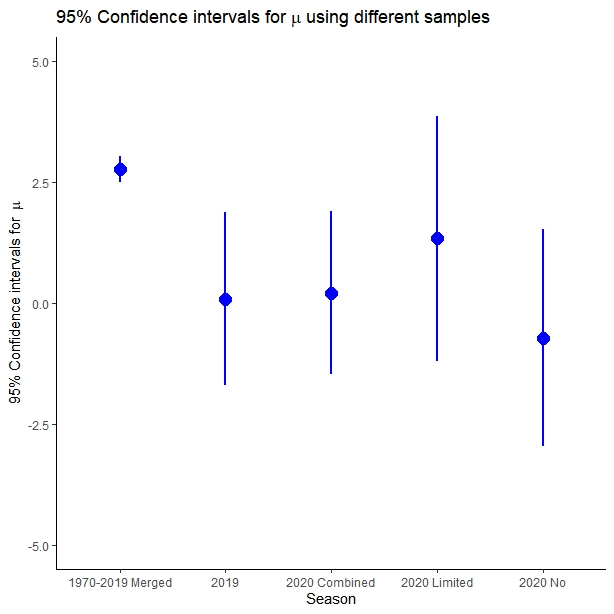}
  \caption{This figure shows the 95\% confidence intervals of the mean of the difference of home team's and away team's points using different observations. From the left to right, we can see the corresponding confidence intervals, where all games from the \textit{merged 1970-2019 sample} were considered, all games from Season 2019 were considered, all games from Season 2020 were considered, games from Season 2020 with limited attendance were considered, and games from Season 2020 with strictly no spectators were considered.}
  \label{Fig_ConfIntNormmeans}
\end{figure}

As already discussed in Section~\ref{Method:Scores}, we are able to construct confidence intervals for the mean of $D$.
In essence, we fit separately for each sample a normal distribution for $D$, and provide confidence intervals for the $\mu$ parameters of the normal distributions.
Corresponding results can be seen in Figure~\ref{Fig_ConfIntNormmeans}. As we can see the 95\% confidence interval for the \textit{merged 1970-2019 sample} is relatively narrow due to the large number of considered games. This 95\% confidence interval for $\mu$ is $[2.51,3.04]$. In this figure we only show the confidence interval for $\mu$ to other four samples, namely to Season 2019 and to the three scenarios of Season 2020. All other confidence intervals do overlap with the confidence interval of the \textit{merged 1970-2019 sample}. Indeed as Figure~\ref{Fig_Diff} shows the highest average of $D$ occurred in Season 1985, and the corresponding confidence interval for $\mu$ is $[2.72,6.55]$. The lowest average of $D$ excluding both Season 2019 and Season 2020 occurred in Season 2006, and the corresponding confidence interval for $\mu$ is $[-0.67,2.77]$. Therefore we cannot argue by comparison of confidence intervals that any year before 2019 was an outlier when compared to the \textit{merged 1970-2019 sample}. As for Figure~\ref{Fig_ConfIntNormmeans}, the confidence intervals are wider for the four latter samples.
Based on the comparison of confidence intervals we can assert that Season 2019 and the entire Season 2020 is significantly different from earlier seasons. On the other hand, when only considering games where some limited audience were allowed the previous argument no longer holds. The average difference between home and away scores considering games from Season 2020 with some limited audience is 1.336, that is no longer a clear outlier when comparing with the values in Figure~\ref{Fig_Diff}. Also the corresponding 95\% confidence interval does overlap with the 95\% confidence interval of the \textit{merged 1970-2019 sample}. Finally, the 95\% confidence interval of Season 2020 with no spectators again does not overlap with the confidence interval of the \textit{merged 1970-2019 sample}. The average of $D$ considering these games is -0.714, being the lowest and only negative value from all samples. On top of this, the corresponding confidence interval of Season 2020 with no audience also has the lowest boundary points. 

Regarding the two sample $t$-test, we again compared different samples to the \textit{merged 1970-2019 sample}. As for the actual $T$ statistic values, we got $T=3.00$ with \textit{Scenario 1} of Season 2020, $T=1.11$ with \textit{Scenario 2} of Season 2020 and $T=2.92$ with \textit{Scenario 2} of Season 2020.
The actual $\text{t}_{1-\frac{\alpha}{2},\nu}$ values are very close for different possible $\nu$-s at a fixed $\alpha$ level.
For $\alpha=95\%$: $1.96<\text{t}_{1-\frac{\alpha}{2},\nu}<1.98$, for $\alpha=99\%$: $2.59<\text{t}_{1-\frac{\alpha}{2},\nu}<2.61$ hold for all $\nu$ values we can consider by comparing a sample to the \textit{merged 1970-2019 sample}.
Therefore even at 99\% confidence level we can reject that the mean of $D$ of the \textit{merged 1970-2019 sample} is the same as for the sample of all games of Season 2020 or for the sample of games of Season 2020 with no audience. Regarding the games of Season 2020 with some limited audience we cannot make the same rejection even at 95\% confidence level.

We again also calculated the two sample $t$-test values as a comparison to the \textit{merged 1970-2019 sample} by choosing those seasons with the help of Figure~\ref{Fig_Diff} that have the lowest or highest average $D$ values. We got $T=-1.88$ for Season 1985, $T=1.94$ for Season 2006 and $T=2.90$ for Season 2019. This means that for the games of Season 2019 even at $99\%$ confidence level we can reject the corresponding null hypothesis such that the mean of $D$ of Season 2019 is the same as the mean of $D$ of the \textit{merged 1970-2019 sample}.
For all other seasons even at $95\%$ confidence level we cannot reject the null hypothesis that the mean of $D$'s are the same.

To sum up, we can argue that Season 2019, \textit{Scenario 1} of Season 2020 and \textit{Scenario 3} of Season 2020 are outliers in terms of difference of scores achieved by the two teams compared to earlier seasons.  Based on the confidence intervals and the $T$-test values we can also argue that the games with no audience in Season 2020 showed the most significantly different behaviour in terms of points scored by the two teams. Home teams underperformed in terms of scored points compared to visiting teams by the largest margin when no spectators were allowed to be present at the stadium. Even though Season 2019 is significantly different in terms of achieved scores, it is not significantly different in terms of home team wins percentage. The only samples that are substantially different using both criteria belong to Season 2020.

\section{Conclusions}

In this paper we investigated the effect of spectators on the game outcome via NFL games.
Results showed that the home team performance during Season 2020 significantly declined compared to previous normal seasons.
In fact, the lowest proportion of home team wins is observed for Season 2020 out of the last 51 entire NFL seasons. 
We also investigated the effect of having some fans at site, and our study reveals that when allowing only up to 25\% of spectators the home team performance is no longer significantly different compared to normal circumstances.
These findings suggest that the only relevant causes that explain general home team advantage are crowd effects, referee bias and psychological factors. These causes already become relevant when stadiums are only partially loaded.
Therefore we claim that all potential causes of home advantage for American football are driven by psychological concepts.
We supported this claim by examining game outcomes and also scored points by the two teams in games.

Future studies can investigate these findings considering different sports. The methodologies of this paper can be applied with minor modifications for other team sports. 


\bibliography{sample.bib}

\end{document}